\documentclass[nocompress]{spie}  

\usepackage{amsmath,amsfonts,amssymb}
\usepackage{graphicx}
\usepackage{xcolor}
\usepackage{natbib}
\usepackage{caption}
\usepackage{subcaption}

\def\apjl{ApJ}%
\def\aap{A\&A}%
\usepackage{amsmath,amsfonts,amssymb}
\usepackage{graphicx}
\usepackage{xcolor}
\usepackage{natbib}
\def\apjl{ApJ}%
\def\aap{A\&A}%
\usepackage[colorlinks=true, allcolors=blue]{hyperref}

\title{NPF update: Light-weight mirror development in Chile}

\author[a,b,d]{A. Bayo}
\author[b]{P. Mardones}
\author[a,b]{S. Castillo}
\author[c,d]{G. Hamilton}
\author[a,b]{C. Lobos}
\author[b,c,d]{L. Pedrero}
\author[b,c,d]{C. Rozas}
\author[b,c]{N. Soto}
\author[b,c,d]{H. Hakobyan}
\author[b,c,d]{C. Garc\'ia}
\author[b,c]{M. R. Schreiber}
\author[b,c,d]{W. Brooks}
\author[a,b,e]{S. Z\'u\~niga-Fern\'andez}
\affil[a]{Instituto  de  F\'isica  y  Astronom\'ia,  Facultad  de  Ciencias,  Universidad de Valpara\'iso, Chile}
\affil[b]{N\'ucleo Milenio de Formaci\'on Planetaria - NPF, Valpara\'iso, Chile}
\affil[c]{Universidad T\'ecnica Federico Santa Mar\'ia}
\affil[d]{Centro Cient\'ifico Tecnol\'ogico de Valpara\'iso, CCTVal}
\affil[e]{European Southern Observatory, Santiago de Chile, Chile}

\authorinfo{Further author information: (Send correspondence to A.B.)\\A.B.: E-mail: abayo@npf.cl, Telephone: +56 32 250 8311}

\begin{document} 
\maketitle

\begin{abstract}
Planet Formation research is blooming in an era where we are moving from speaking about ``protoplanetary disks" to ``planet forming disks" \citep{Andrews18}. However, this transition is still motivated by indirect (but convincing) hints. Up to date, the direct detection of planets ``in the making" remains elusive with the remarkable exception of PDS\,70\,b and c \citep{Keppler18, Mueller18, Haffert19}. The scarcity of detections is attributable to technical challenges, and even for the rare jewels that we can detect, characterization (resolving their hill spheres) is unachievable. The next step in this direction demands from near to mid-infrared interferometry to jump from $\sim$100\,m baselines to $\sim$1\,km, and from very few telescopes (two to six) to 20 or more (PFI like concepts, \cite{Monnier18}). This transition needs for more affordable near to mid-infrared telescopes to be designed. Since the driving cost for such telescopes resides on the primary mirror, in particular scaling with its diameter and weight, our approach to tackle this problem relies on the production of low-cost light mirrors.
\end{abstract}


\section{INTRODUCTION}
\label{sec:intro}  

Planet formation is arguably one of the hottest topics in observational (and theoretical) astronomy. Thousands of fully formed planetary systems have been confirmed and small and large scales structures, hinting probably the presence of young planets, seem ubiquitous among protoplanetary disks (e.g. \cite{Andrews18}).
Despite all these exciting results, the direct detection of planets ``in the making" remains elusive with the remarkable exception of two point sources: PDS\,70\,b and c \citep{Keppler18, Mueller18, Haffert19}. The scarcity of detections is attributable to technical challenges, and even for the rare jewels that we can detect, resolving their hill spheres to understand the key aspects of their formation is an unachievable endeavour.

For this next step to happen, given the extremely small scale and complicated environment involved, near to mid-infrared interferometers are the only viable option, but these have to jump from $\sim$100\,m baselines to $\sim$1\,km, and from very few telescopes to 20 or more (e.g PFI like concepts, \cite{Monnier18}).

The list of challenges that this leap has to face is not small, and even if the long-base baselines technical ones were to be overcome in the near future (see for example \cite{Bourdarot20} for a promising avenue), the cost of 20 or more near-to-mid (NIR--MIR) infrared three to four or even eight meter class telescopes remains a major possible hurdle for such facility to become a reality.

Since 2017, a group of astronomers, physicists and engineers from Universidad de Valpara\'iso and Universidad T\'ecnica Federico Santa Mar\'ia (both in Valpara\'iso, Chile), has been working on the development of solutions to alleviate the cost of NIR--MIR telescopes (see \cite{zuniga18} for a first report on our activities).

In this wavelength domain (3-10\,$\mu$m), the cost of the telescope is intimately linked to the diameter of its primary mirror, and even if an exponential relation does not hold for segmented primaries, the primary mirror (and its weight) is still the main driver of the cost of the facility.

Our approach to tackle this problem is to use Carbon Fiber Reinforced Polymer (CFRP) as the base to build segmented primaries taking full advantage of the lightness of this material, its strength, its relatively small expansion coefficient, and the opportunity that it brings to ``serialize" the process of astronomical mirror production.

\begin{figure}[!ht]
         \centering
         \includegraphics [width=\textwidth]{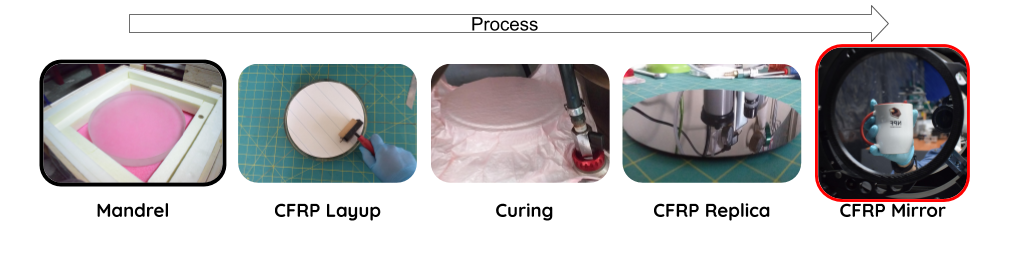}
         \caption{Main steps of the CFRP Mirror replication method. For more details on the lay-up step and the control through the process see Castillo et al. and Soto et al.}
         \label{fig:CFRP Mirror replication method}
\end{figure}

In short, the process of building a CFRP mirror is displayed in Fig.~\ref{fig:CFRP Mirror replication method} and consists of the following steps:

\begin{enumerate}
\item Building of a convex ``mold" (mandrel) of the opposite curvature than that of the final mirror (but with similar or better surface roughness than the goal one).
\item Laying-up and replicating such surface with some kind of reinforced carbon fiber sheet/gel/etc.
\item Curing the copy / replica.
\item Detaching the replica from the mandrel and ``double" coating to maximize reflectivity and protect the surface.
\end{enumerate}

Once a reflective high quality replica is produced, it needs to be integrated with the rest of the telescope through a suitable support cell.

Of course, every one of the previously mentioned steps is meaningless without a set of measurements and quality control processes.

Since the previous SPIE edition we have made substantial progress in all these areas and we are submitting accompanying papers focused on lay-up, and control and metrology (see Castillo et al. and Soto et al., respectively). This paper ``simply" aims at reporting on the general progress of the project, suggestions for support cells being designed for light mirrors, and rough estimates of the cost model for our replicas.

\section{Mandrels}

The performance of different materials for mandrels has been tested in the past two years. Our range covered from the previously used steel mandrel that induced large scale deformations and offered itself non-optimal roughness, to different internationally and locally produced glass and glass-ceramics recipes. In Table~\ref{tab:mandrels} we show a summary of the materials analyzed and conclude that the best results ($\sim$10 \,nm roughness) are systematically obtained for silicate or even regular crown glass.

\begin{table}
\caption{Summary of the performance as mandrels for different materials explored in our laboratory.}
\label{tab:mandrels}
\label{tab:grid}
\centering
\begin{tabular}{lcccccccc}
\hline\hline
   & \multicolumn{8}{c}{Glass} \\
\hline
Kind & Silica & Crown & Pyrex & BK7 & UNK & Pyrex & Crown & Crown \\

\hline
Size (mm) & 150 & 200 & 200  & 70  & 190  & 500  & 500  & 250 \\
\# replicas & 28 & 20 & 54 & 10 & 19 & 5 & 2 & 14 \\
\# replicas \\
before damage & 20 & & 44 & 10 & 19 &  &2 & 13 \\
Thermal exp. \\
coeff. (K$^{-1}$) & 0.5 & 8 & 3.25 & 3.3 & & 3.25 & 8 & 8  \\
Roughness (nm) & 10  & 20  & 12  & 10  & 10  & 13  & 10  & 25 \\
Shape  & spheric & spheric & spheric & flat & spheric & spheric & spheric & spheric \\
Hardness \\
(Mohs scale) & 7 & 7 & 7 & 7 &  & 7 & 7 &7\\
\hline
\end{tabular}

\vspace{0.5cm}
\begin{tabular}{lccccccc}
\hline\hline
   &  \multicolumn{2}{c}{Ceramics} & \multicolumn{2}{c}{Metal} & Marble & Gypsum & CFRP \\
\hline
Kind &  & glass-ceramic & aluminium & steel &  & Type 4 & \\

\hline
Size (mm) & 100 & 50 & 50x50 &500 & 75 & 200& 10\\
\# replicas & 0 & 1 & 15 & 20 & 1 & 1 & 0\\
\# replicas \\
before damage & N/A & 1 &N/A& N/A& 1 & 1& N/A \\
Thermal exp. \\
coeff. (K$^{-1}$)  &2.5&2.5& 23 & 12 & 7.2 & 18 & -0.8 \\
Roughness (nm) &   & $\sim$100 &  &  & 70 & 120 &  \\ 
Shape  & flat & spheric & flat & spheric & spheric & spheric & flat \\
Hardness \\
(Mohs scale) &   &  &  &  &  & 2 & \\
\hline
\end{tabular}
\end{table}

The largest optical-quality mandrel produced by us so far has a diameter of barely 30\,cms, but a 0.5\,m one is being polished in our lab with a current measured roughness of a very few tens of nm (not included in Table~\ref{tab:mandrels}).

A very important advantage of CFRP replicas is precisely its replicability, but, obviously, the processes involved infer stress on the mandrels. We have been monitoring any deterioration of the mandrels and estimate that several tens of replicas can be obtained from a single mandrel before it deteriorates (see Table~\ref{tab:mandrels}). A current line of research in this aspect deals with mandrel reparation and we plan on reporting on it during 2021.

\section{Lay-up}

As previously mentioned, the goal to ``obtain a negative copy" of the mandrel with a composite material (such as CFRP) is to obtain a high-fidelity replica. Such replica could / should exhibit the same characteristics in shape (although opposite curvature) and roughness than the mandrel. Unfortunately, print-through and other kinds of ``pattern transfer" from the individual carbon fibers present in composite materials can result in a ``waviness signature" ($\sim$100\,nm periodic patterns) that can seriously harm the utility of such replicas for interferometry. 

This fiber print-through problem is not an unknown one and several approaches already exist in the literature to mitigate its effects. The most common one consist of incorporating additional layers of epoxy or even metals that will ``smear out" the unwanted pattern (see for example \cite{Steeves14}). 

During the past two years we have obtained very promising results to circumvent this problem by not only applying additional epoxy layers, but tackling the problem early on, from the lay-up strategy. Regarding the latter, for further details check Castillo et al. (in this same volume), but in short, the proposed method basically consists on changing the orientation of a number of layers at the beginning of the lay-up, maintaining three simple considerations: symmetry, balance and quasi-isotropy. 

These changes in lay-up resulted in a significant improvement at high and mid-spatial frequencies, usually corresponding to roughness and waviness (see Castillo et al. for details).


\section{Curing}

Temperature and pressure control are vital in handling the viscosity of the epoxy either present in pre-impregnated CFRP or used independently in the replica process (and to reach the glass transition temperature of the resin). 

We have been working on optimizing a pure baking and autoclave-curing sequence to maximise the fidelity of the replica at all spatial frequencies. Unfortunately, the autoclave that provided the best results in terms of overall shape is a ``wet" one that resulted in humidity pockets developing in between layers and therefore yielding sub-optimal roughness performance.

\begin{figure}[!ht]
     \centering
     \begin{subfigure}[b]{0.475\textwidth}
         \centering
         \includegraphics[width=\textwidth]{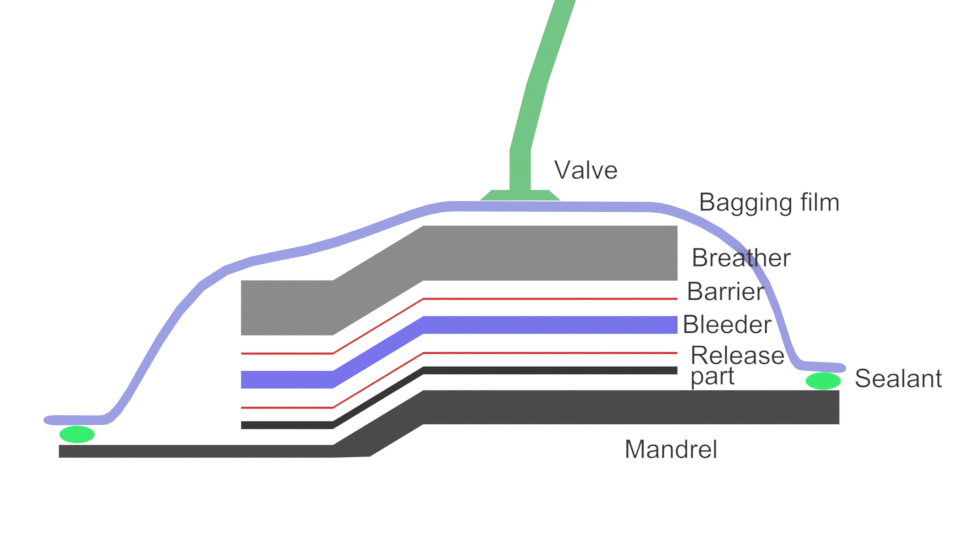}
         \caption{Sketch of the different layers present in the vacuum bag during the first curing cycle in the oven.}
     \end{subfigure}
     \hfill
     \begin{subfigure}[b]{0.47\textwidth}
         \centering
         \includegraphics[width=\textwidth]{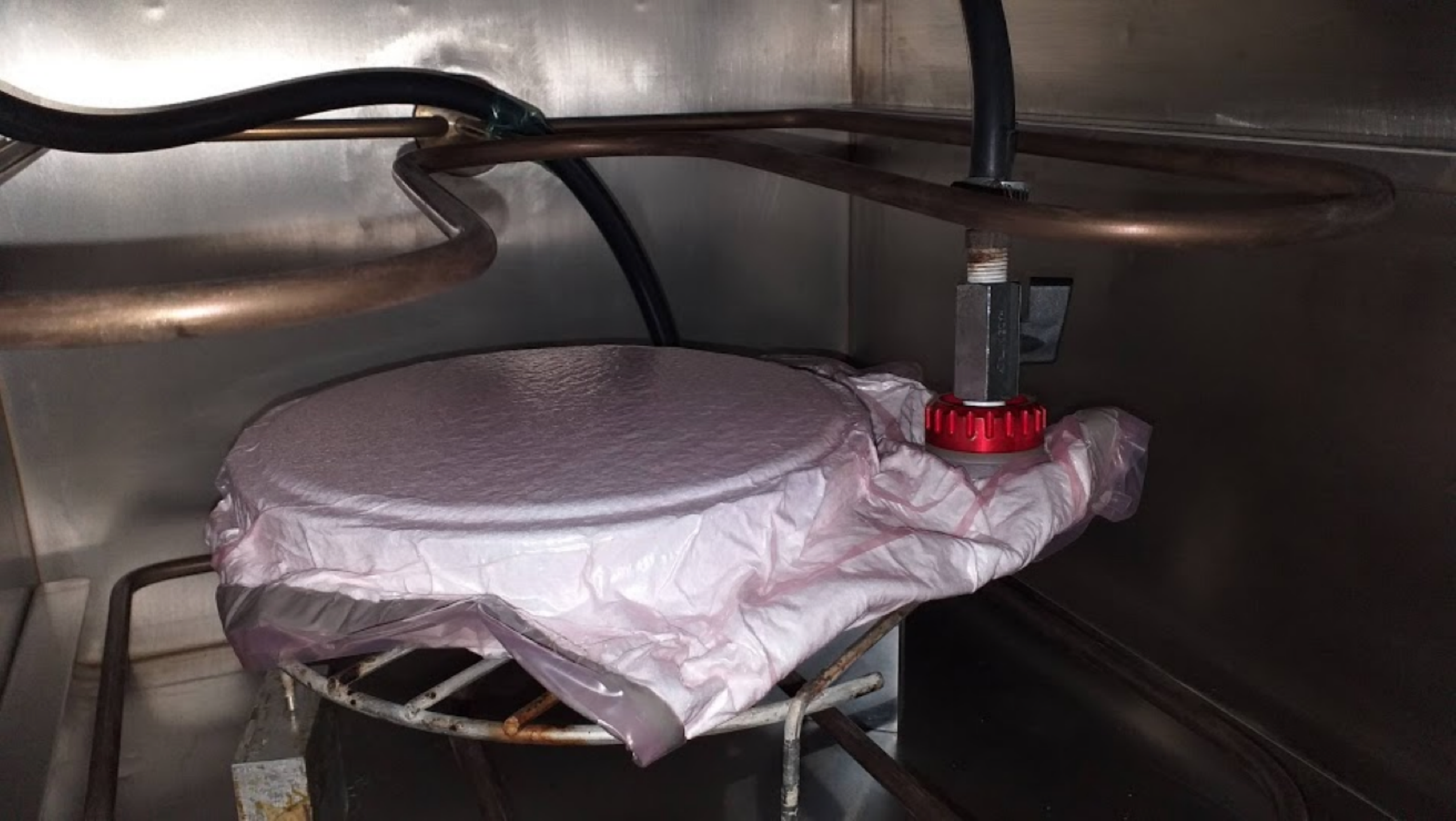}
         \caption{Picture of a replica in the process of baking in the oven.}
     \end{subfigure}
        \vspace{0.25cm}
        \caption{Example of step one on our two-step curing process.}
        \label{fig:oven}
\end{figure}

While we are in the process of receiving a large (1m in diameter capacity) autoclave, we have proceeded with a two step curing methodology: a higher temperature one where the pre-impregnated CFRP is cured (see Fig.~\ref{fig:oven} for a sketch of the layering process involved in the ``baking"), and a secondary room temperature curing where additional layers of epoxy are added and imperfections at the smallest and intermediate frequencies are corrected.

This two-step approach provides virtually one to one copies from the mandrels in terms of roughness and waviness, however, first order aberrations are still present. Our preliminary approach to solve this problem is prototyping a simple ``one-time-active" support cell that serves also as a device to integrate the mirror in different optical setups (see section~\ref{sec:cell}).

\section{Unmolding and coating}

Different release agents and application procedures have been tested through the last two years in order to find the right balance between optimal unmolding, residual-free surfaces, and minimum damage to the mandrel. 

As mentioned before, the practically negligible small scale surface degradation in the whole replicating process and the large number of replicas that we obtain from a given mandrel before considering reparation (see Table~\ref{tab:mandrels}), suggest that we have achieved that balance (see Soto et al. for details on our quality control and metrology processes).

Regarding coating, we are following two paths: evaporation and sputtering. Although the latter should provide longer lasting results (based on glass coating experience), the life-time / deterioration of the reflectivity of the CFRP replicas (in particular on top of the additional resin layers) is still a matter of study. 

\begin{figure}[!ht]
     \centering
     \begin{subfigure}[b]{0.5\textwidth}
         \centering
         \includegraphics[width=\textwidth]{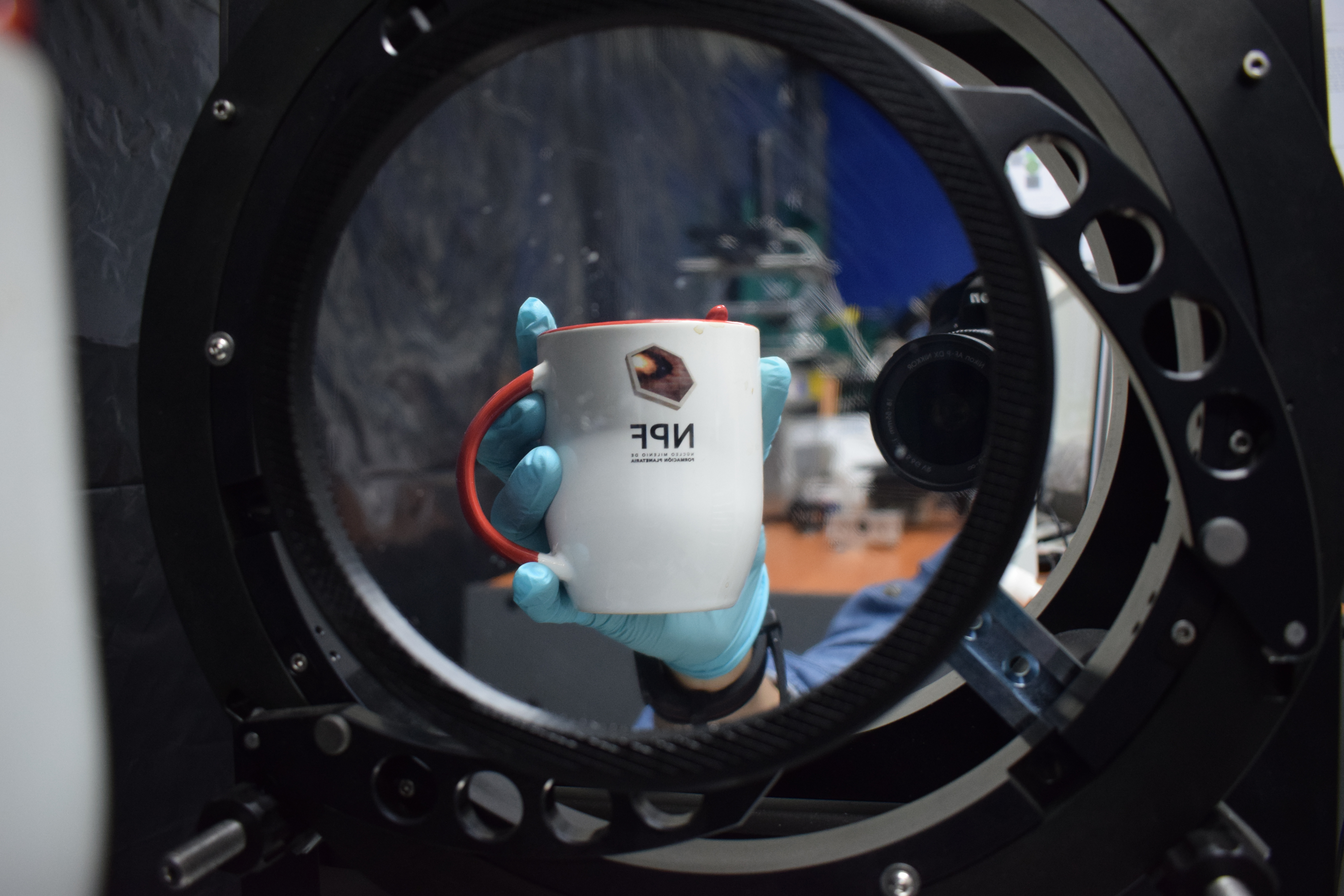}
         \caption{First CFRP replica with successful aluminium coating in our lab.}
     \end{subfigure}
     \hfill
     \begin{subfigure}[b]{0.4\textwidth}
         \centering
         \includegraphics[width=\textwidth]{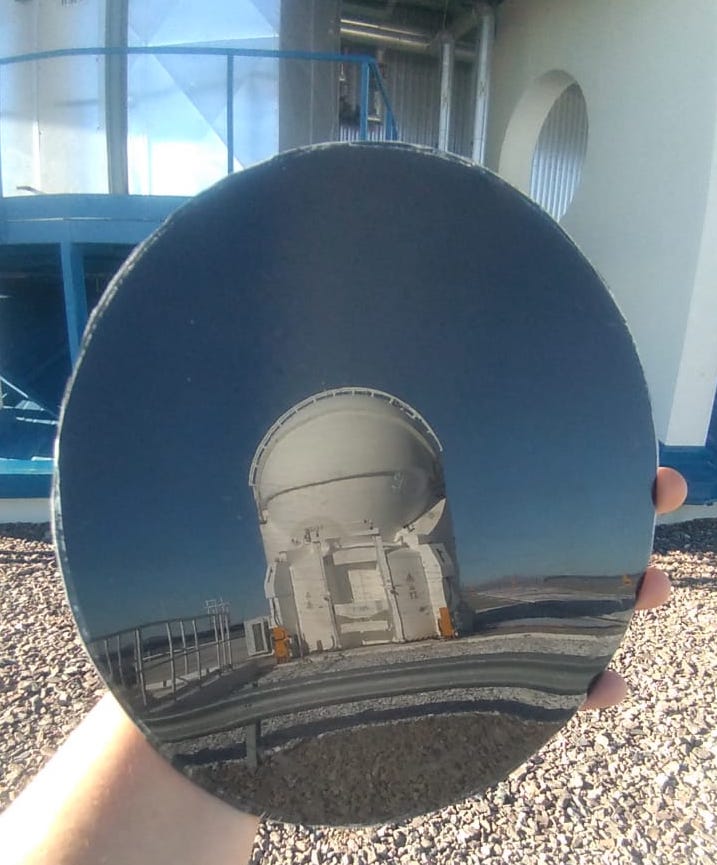}
         \caption{Metallic uncoated replica transported to Paranal Observatory.}
     \end{subfigure}
        \vspace{0.25cm}
        \caption{Aluminium coated and uncoated CFRP replicas.}
        \label{fig:coating}
\end{figure}

So far, we have experimented with copper (that poses adherence problems, both, on replicas with layers of extra resin and those without) and aluminium on $\sim$20\,cm replicas with single target strategy. Aluminium presented no adherence issues and the usual thickness of 100\,nm yielded good reflectivity, although, as mentioned before, long term changes in reflectivity due to deterioration (mostly oxidation) of the surface have not yet been studied. By the end of 2019 we took some of our first coated replica, and some uncoated (see Fig.~\ref{fig:coating}) to ESO's Paranal Observatory, to be exposed to the dust and wind of the Atacama desert, and we plan on studying their deterioration during 2021. 

The purpose of this experiment is not only to characterize the adherence of dust to the coated surfaces and the possibility to clean it without affecting reflectivity, but also to monitor any changes on uncoated / unprotected surfaces that may arise from extreme changes in humidity, for instance due to pockets of humidity locked within the layers in our lab (that is located in a very high humidity region). Our preliminary assessments in Paranal regarding surface roughness of the uncoated replica is that there is not significant deterioration.

Even though we have not been able to characterize the deterioration of the coating, previous experience with glass suggest that a protective layer is key in ensuring long-lasting highly reflective surfaces, therefore we have also experimented with two target sputtering coating. 

In this two targets approach, aluminium was still the material of choice for the reflective layer and silicon dioxide that of the protective layer. The choice of SiO$_2$ as protective layer was motivated by the literature, given its good characteristics for the near-infrared (further tests for mid-infrared absorption, etc need to be carried out still). Different thicknesses for the protective layers are still being tested with the plan to compare reflectivity over time for replicas with 50, 100 and 150\, nm of SiO$_2$ (over our ``standard" 100\,nm layer of Al).

Our goal for 2021 is to commission a 50\,cm coating machine and repair the 2 meter sputtering machine available to us, as well as to advance on two target coating characterization.

\section{Support cells}
\label{sec:cell}

As explained before, achieving a perfect copy of the mandrel at all spatial frequencies is something still out of our reach. However, we have found ways to obtain excellent results concerning waviness and surface roughness, with the ``only" pending issue being that of first order aberrations, mostly astigmatism directly linked to the orientation of the fibers in the ``last" layer of CFRP sheet.

While we are still tackling this problem from the manufacturing side (via additional optimization of the lay-up process an ``a posteriori" resign layer additions), we are also developing a cell system with two main goals in mind: to provide support to the CFRP mirror (and therefore serve as a key element for the integration of our mirrors in different optical setups) and to be able to correct any first order aberration. This cell is mainly composed of 3 parts:

\begin{enumerate}
\item A ring-type piece that acts as a boundary condition, keeping the perimeter of the CFRP mirror correctly shaped.
\item A body with a thread that fits with the ring-type piece, keeping the CFRP mirror in between those and also acting as a base for the actuators.
\item An array of threaded bolts with spherical heads as actuators.
\end{enumerate} 

\begin{figure}[h]
     \centering
         \centering
         \includegraphics[width=0.6\textwidth]{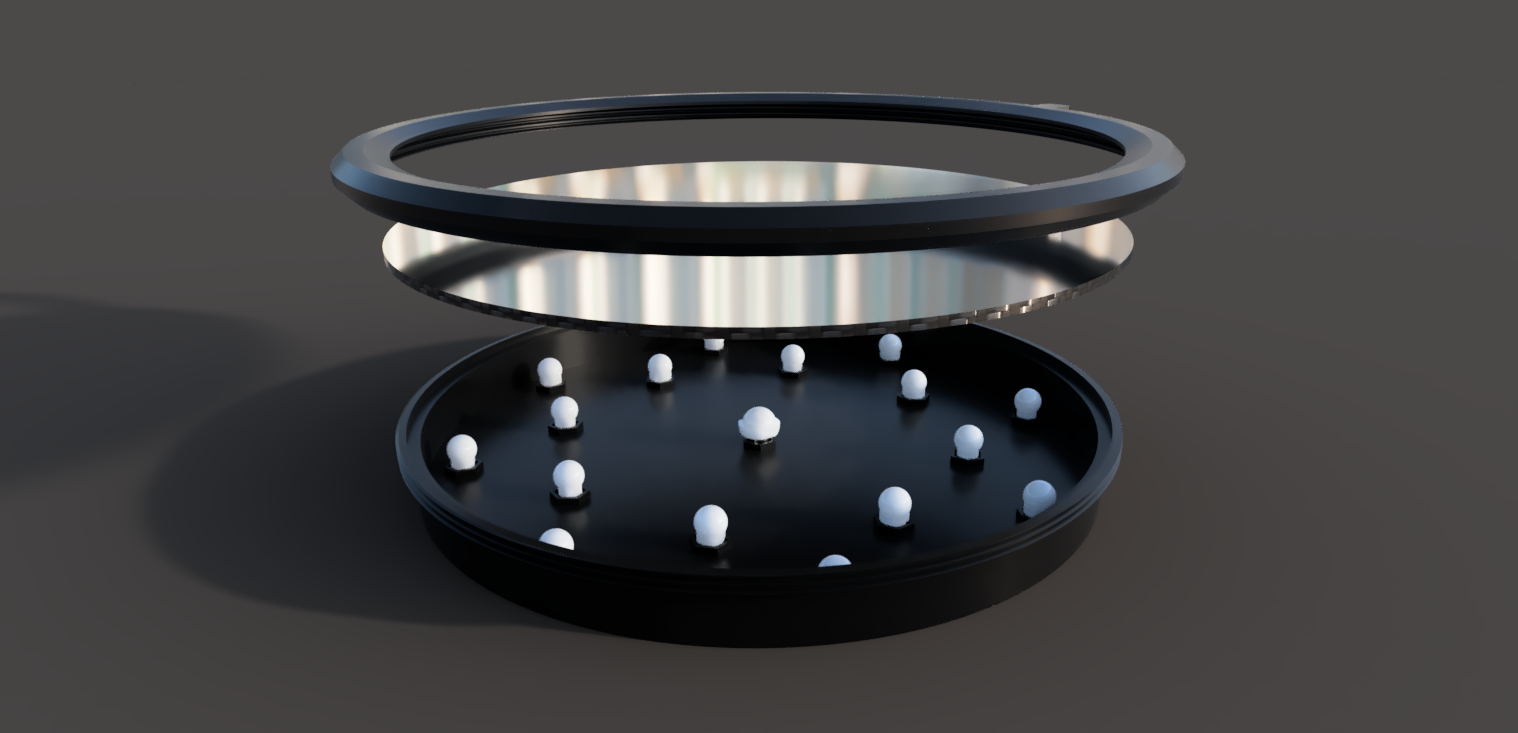}
         \vspace{0.2cm}
         \caption{3D CAD model of a third generation prototype of a CFRP mirror support cell.}
         \label{fig:mirror_cell}
\end{figure}

We have so far produced three different versions of this supporting cell 3D printing with different materials to characterize accuracy of the correction and stability of the latter. This characterization is at its early stages and therefore we cannot offer proper quantification of the image quality improvement yet.

\section{Cost model}

A main motivation to carry out this research on CFRP mirrors is the hypothesis that these mirrors will prove to be of similar quality than the ``traditional" glass-based one but with a much lower cost.

Even though the size of the replicas that we are producing are still of very moderate size, we performed the exercise to build a parametric cost model to test weather the possible market prize will go in the right direction.

Accounting for human resources, machinery investment, mandrel acquisition, etc., our model places the cost of a $\sim$20\,cm replica ``ready to go" well below 450 USD; and a rough estimate for a mount that could handle such a light-weight primary (barely 200grams) yields a figure of less than 750 USD.

In principle, there is no reason to assume anything but linearity regarding the scalability in the cost of the replica process itself, and if that is the case, CFRP mirrors can definitely be the change in paradigm needed for facilities like PFI to become a reality.

\acknowledgments 
 
All the authors acknowledge financial support from Iniciativa Cient\'ifica Milenio v\'ia N\'ucleo Milenio de Formaci\'on Planetaria. A.B acknowledges support from FONDECYT grant 1190748, A.B. and N. S. acknowledges support from ESO Comit\'e-Mixto and QUIMAL funding agencies. M.R.S., S.C and C.L acknowledge support from the ALMA-CONICYT fund. S.Z-F acknowledges financial support from the European Southern Observatory via its studentship program and ANID via PFCHA/Doctorado Nacional/2018-21181044.

\bibliography{report} 

\begin{thebibliography}{}
\makeatletter
\relax
\def\mn@urlcharsother{\let\do\@makeother \do\$\do\&\do\#\do\^\do\_\do\%\do\~}
\def\mn@doi{\begingroup\mn@urlcharsother \@ifnextchar [ {\mn@doi@}
  {\mn@doi@[]}}
\def\mn@doi@[#1]#2{\def\@tempa{#1}\ifx\@tempa\@empty \href
  {http://dx.doi.org/#2} {doi:#2}\else \href {http://dx.doi.org/#2} {#1}\fi
  \endgroup}
\def\mn@eprint#1#2{\mn@eprint@#1:#2::\@nil}
\def\mn@eprint@arXiv#1{\href {http://arxiv.org/abs/#1} {{\tt arXiv:#1}}}
\def\mn@eprint@dblp#1{\href {http://dblp.uni-trier.de/rec/bibtex/#1.xml}
  {dblp:#1}}
\def\mn@eprint@#1:#2:#3:#4\@nil{\def\@tempa {#1}\def\@tempb {#2}\def\@tempc
  {#3}\ifx \@tempc \@empty \let \@tempc \@tempb \let \@tempb \@tempa \fi \ifx
  \@tempb \@empty \def\@tempb {arXiv}\fi \@ifundefined
  {mn@eprint@\@tempb}{\@tempb:\@tempc}{\expandafter \expandafter \csname
  mn@eprint@\@tempb\endcsname \expandafter{\@tempc}}}

\bibitem[\protect\citeauthoryear{{Andrews} et~al.,}{{Andrews}
  et~al.}{2018}]{Andrews18}
{Andrews} S.~M.,  et~al., 2018, \mn@doi [\apjl] {10.3847/2041-8213/aaf741},
  \href {https://ui.adsabs.harvard.edu/abs/2018ApJ...869L..41A} {869, L41}

\bibitem[\protect\citeauthoryear{{Bourdarot}, {Guillet de Chatellus}  \&
  {Berger}}{{Bourdarot} et~al.}{2020}]{Bourdarot20}
{Bourdarot} G.,  {Guillet de Chatellus} H.,   {Berger} J.~P.,  2020, \mn@doi
  [\aap] {10.1051/0004-6361/201937368}, \href
  {https://ui.adsabs.harvard.edu/abs/2020A&A...639A..53B} {639, A53}

\bibitem[\protect\citeauthoryear{{Haffert}, {Bohn}, {de Boer}, {Snellen},
  {Brinchmann}, {Girard}, {Keller}  \& {Bacon}}{{Haffert}
  et~al.}{2019}]{Haffert19}
{Haffert} S.~Y.,  {Bohn} A.~J.,  {de Boer} J.,  {Snellen} I.~A.~G.,
  {Brinchmann} J.,  {Girard} J.~H.,  {Keller} C.~U.,   {Bacon} R.,  2019,
  \mn@doi [Nature Astronomy] {10.1038/s41550-019-0780-5}, \href
  {https://ui.adsabs.harvard.edu/abs/2019NatAs...3..749H} {3, 749}

\bibitem[\protect\citeauthoryear{{Keppler} et~al.,}{{Keppler}
  et~al.}{2018}]{Keppler18}
{Keppler} M.,  et~al., 2018, \mn@doi [\aap] {10.1051/0004-6361/201832957},
  \href {https://ui.adsabs.harvard.edu/abs/2018A&A...617A..44K} {617, A44}

\bibitem[\protect\citeauthoryear{{Monnier} et~al.,}{{Monnier}
  et~al.}{2018}]{Monnier18}
{Monnier} J.~D.,  et~al., 2018, \mn@doi [Experimental Astronomy]
  {10.1007/s10686-018-9594-1}, \href
  {https://ui.adsabs.harvard.edu/abs/2018ExA....46..517M} {46, 517}

\bibitem[\protect\citeauthoryear{{M{\"u}ller} et~al.,}{{M{\"u}ller}
  et~al.}{2018}]{Mueller18}
{M{\"u}ller} A.,  et~al., 2018, \mn@doi [\aap] {10.1051/0004-6361/201833584},
  \href {https://ui.adsabs.harvard.edu/abs/2018A&A...617L...2M} {617, L2}

\bibitem[\protect\citeauthoryear{{Steeves}, {Laslandes}, {Pellegrino},
  {Redding}, {Bradford}, {Wallace}  \& {Barbee}}{{Steeves}
  et~al.}{2014}]{Steeves14}
{Steeves} J.,  {Laslandes} M.,  {Pellegrino} S.,  {Redding} D.,  {Bradford}
  S.~C.,  {Wallace} J.~K.,   {Barbee} T.,  2014, in {Navarro} R.,  {Cunningham}
  C.~R.,   {Barto} A.~A.,  eds,  Society of Photo-Optical Instrumentation
  Engineers (SPIE) Conference Series Vol. 9151, Advances in Optical and
  Mechanical Technologies for Telescopes and Instrumentation. p. 915105,
  \mn@doi{10.1117/12.2056560}

\bibitem[\protect\citeauthoryear{{Z{\'u}{\~n}iga-Fern{\'a}ndez}
  et~al.,}{{Z{\'u}{\~n}iga-Fern{\'a}ndez} et~al.}{2018}]{zuniga18}
{Z{\'u}{\~n}iga-Fern{\'a}ndez} S.,  et~al., 2018, in {Marshall} H.~K.,
  {Spyromilio} J.,  eds,  Society of Photo-Optical Instrumentation Engineers
  (SPIE) Conference Series Vol. 10700, Ground-based and Airborne Telescopes
  VII. p. 107003X (\mn@eprint {arXiv} {1807.06668}),
  \mn@doi{10.1117/12.2313983}

\makeatother
\end{thebibliography}
\bibliographystyle{mnras} 

\end{document}